# Early warnings of COVID-19 outbreaks across Europe from social media?

Milena Lopreite[1], Pietro Panzarasa[2*], Michelangelo Puliga[3], Massimo Riccaboni[4]

[1]Department of Economics, Statistics and Finance, University of Calabria (Italy); Institute of Management, Scuola Superiore Sant'Anna, Pisa (Italy).

[2]School of Business and Management, Queen Mary University of London, London, United Kingdom.

[3]IMT School for Advanced Studies, Lucca (Italy); Institute of Management, Scuola Superiore Sant'Anna, Pisa (Italy); Linkalab Computational Laboratory, Cagliari (Italy).

[4]IMT School for Advanced Studies, Lucca (Italy).

**Abstract:** We analyze data from Twitter to uncover early-warning signals of COVID-19 outbreaks in Europe in the winter season 2019-2020, before the first public announcements of local sources of infection were made. We show evidence that unexpected levels of concerns about cases of pneumonia were raised across a number of European countries. Whistleblowing came primarily from the geographical regions that eventually turned out to be the key breeding grounds for infections. These findings point to the urgency of setting up an integrated digital surveillance system in which social media can help geo-localize chains of contagion that would otherwise proliferate almost completely undetected.

**Introduction**

Public health surveillance plays a critical role in helping national governments to monitor the emergence of infectious diseases, promptly identify a state of emergence, and propose effective measures to curb an outbreak [1]. Since January 2020, when the severe acute respiratory syndrome–coronavirus 2 (SARS-CoV-2), which causes the Coronavirus disease 2019 (COVID-19), began to spread from China to Europe and the United States, criticism has intensified over the ways in which public health authorities across many countries managed to face the urgency of the threat and devise appropriate mitigation strategies. Lapses in identifying early-warning signals left many national governments largely blind to the unprecedented scale of a looming public health emergency and unable to spur a no-holds-barred timely defense, with severe consequences in terms of mortality rates [2].

Different surveillance strategies have been used to monitor the spread of a disease, including sentinel surveillance systems, household surveys, laboratory-based surveillance, community-based surveillance practices, wastewater surveillance, and the Integrated Disease Surveillance and Response (IDSR) framework [3, 4]. More recently, social media have begun to gain a prominent role as complementary surveillance systems for monitoring epidemics and informing the judgements and decisions of public health officials and experts [5]. For instance, recent work has relied upon multiple digital data streams to uncover early-warning indicators of variations in state-level US COVID-19 activity that may facilitate the detection of impending COVID-19 outbreaks [6]. Leveraging social media to detect early-warning signals of an upcoming pandemic is indeed a good example of epidemiological monitoring [7, 8]. Here, we take a step in this direction, and use social media to show how the general public reacted to emerging epidemic threats by raising anomalous levels of concern on symptoms that are typically associated with COVID-19.

To this end, we have analyzed data from Twitter across a number of European countries to show that unexpected levels of concerns about pneumonia had been raised for several weeks before the first cases of infection were officially announced. Interestingly, we also show that whistleblowing came primarily from the geographical regions that turned out to be the key breeding grounds for infections. Our infodemiological approach to studying the spread of COVID-19 across Europe can help policymakers to better identify, geo-localize and manage chains of infection across national borders and linguistic barriers.

**Results**

On 31 December 2019 the World Health Organization (WHO) was informed about the first "cases of pneumonia of unknown etiology" [9]. This prompted us to rely on pneumonia for detecting early-warning signals of the upcoming pandemic. In particular, we focused on pneumonia for two reasons: (i) pneumonia is the most severe condition induced by COVID-19 [10]; and (ii) the flu season in 2020 was milder than in previous flu seasons [11,12]. We created a unique database including all messages containing the keyword "pneumonia" in the seven most spoken languages of the European Union (i.e., English, German, French, Italian, Spanish, Polish, and Dutch) [13], and posted on Twitter over the period from 1 December 2014 to 1 March 2020. We made a number of adjustments to avoid overestimation of the number of tweets mentioning cases of pneumonia between December 2019 and January 2020 (see details in Methods). In particular, we removed the effects on posting activity of COVID-19-related news that appeared up to 21 January 2020, when COVID-19 became a Class B notifiable disease [12]. Indeed it is reasonable to expect most tweets posted after this date and mentioning pneumonia to be related to the COVID-19 outbreak, even when they did not directly use the word "COVID". For this reason, with regard to tweets posted after 21 January 2020, there would be no obvious way to disambiguate messages concerned with genuine local cases of pneumonia from messages elicited by mass media coverage of the outbreak.

Figure 1a shows the cumulative distribution functions of the normalized number of tweets mentioning the word "pneumonia" in the selected European countries: France, Germany, Italy, the Netherlands, Poland, Spain, UK. To better understand the change in slope exhibited by the curves in the first few weeks of 2020, for each country we conducted a two-sample Kolmogorov-Smirnov (K-S) test of the null hypothesis that the cumulative distributions over two corresponding winter seasons (2018-2019 and 2019-2020) are the same against the alternative hypothesis that they differ. Figure 1b suggests that, with the exception of Germany, the distributions in the two winter seasons are statistically different for all countries: at the 0.10 level of significance for Poland, and at the 0.05 level of significance for the remaining countries (see also Supplementary Table S1 for the details on the specific time periods in which the distributions differ). To check for robustness, we also computed the Anderson-Darling (A-D) test and obtained similar results (Supplementary Fig. S1 and Supplementary Table S2). Finally, we further performed similar robustness checks (i.e., KS and AD tests) by comparing the 2019-2020 winter season with each of the corresponding winter seasons since 2014 (i.e., 2014-2015, 2015-2016, 2016-2017, 2017-2018, and 2018-2019),

and obtained similar findings (Supplementary Fig. S2, Supplementary Fig. S3, Supplementary Table S3 and Supplementary Table S4).

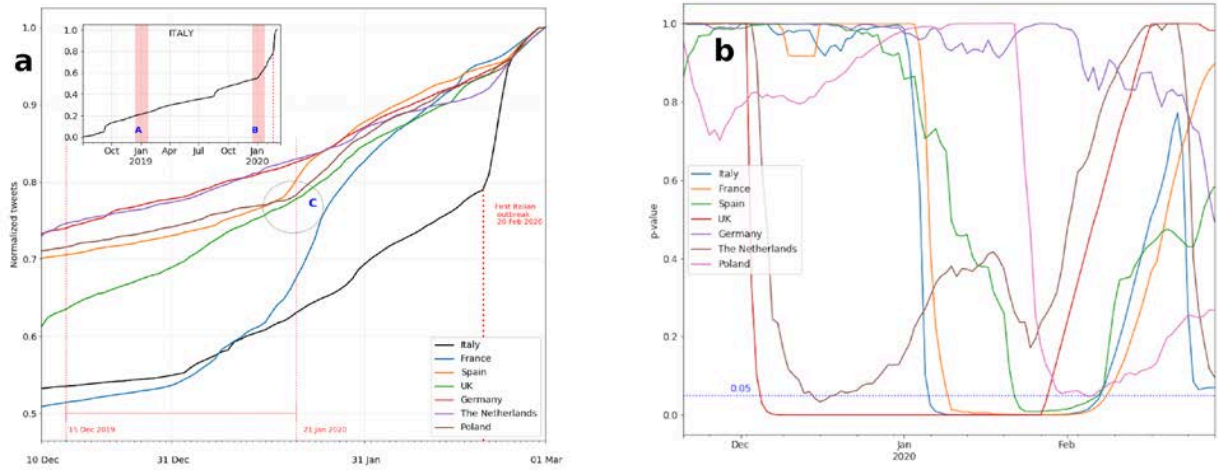

**Fig. 1. Anomalous evolution of pneumonia-related tweets posted across Europe since December 2019.** (a) Cumulative rescaled number of tweets citing pneumonia from 10 December 2019 to 1 March 2020. Inset plot shows the evolution of such tweets posted in Italy from 1 July 2019 to 1 March 2020, and highlights the two winter seasons (shaded bars) used to uncover anomalous spikes of pneumonia-related tweets. (b) Two-sample Kolmogorov-Smirnov test of the difference between cumulative distributions of number of tweets citing pneumonia and posted in the two corresponding winter seasonal periods (2018-2019 and 2019-2020) for each of the 7 European countries. The graph reports the average *p*-values over moving window widths $w \in [50, 70]$ computed with daily frequency.

With the exception of Germany, the series of cumulative mentions of pneumonia unmask unexpected statistically significant variations in public interest in pneumonia-related cases already in January 2020 (Fig. 1a). Interestingly, findings suggest a significant increase in tweets mentioning pneumonia in most of the selected European countries well before the outbreak of COVID-19 was officially reported in the news. In Italy, for example, where the first lockdown measures to contain an emerging threat of endemic COVID-19 infections were introduced on 22 February 2020, the rate of increase in mentions of pneumonia during the first few weeks of 2020 (shaded bar B, inset of Fig. 1a) substantially differs from the rate observed in the same weeks in 2019 (shaded bar A). That is, potentially hidden infection hotspots were identified several weeks before the announcement of the first local source of a COVID-19 infection (20 February, Codogno, Italy). France exhibited a similar pattern, whereas Spain, Poland and the UK witnessed a delay of two weeks (circle C, Fig. 1a). In the Netherlands, after a slow increase subsequent to the COVID-19 outbreak in January 2020, it was only at the end of February 2020 that the curves become steeper. This is likely to reflect local differences in the perception of the disease, as well as differences in the COVID-19 diffusion patterns and infection rates across Europe. From 20 February 2020 the slopes of the curves are likely attributable to a widespread increase in public interest in the pandemic threat across all countries.

We also uncovered variations in the number of users citing pneumonia across European regions between the winter season of 2020 and the corresponding season in the previous year. We obtained the locations of 13,088 users, and identified the European regions that were characterized by anomalous and unexpected surges in pneumonia-related Twitter mentions during the early undetected phases of the COVID-19 outbreak. Figure 2a shows the geographic distribution of unique users discussing pneumonia between 15 December 2019 and 21 January 2020, after filtering out press releases and news accounts. Figure 2b shows the relative increase in the number of such users ($N_U$) between 2020 and the corresponding winter period in 2019 (i.e., $(N_{U, 2020} - N_{U, 2019}) / N_{U, 2019}$). Both maps suggest interesting patterns with respect to the known outbreak evolution: the majority of users discussing cases of pneumonia came precisely from the regions, such as Lombardy, Madrid, Île de France and England, that eventually reported early cases of the COVID-19 contagion (Supplementary Table S5).

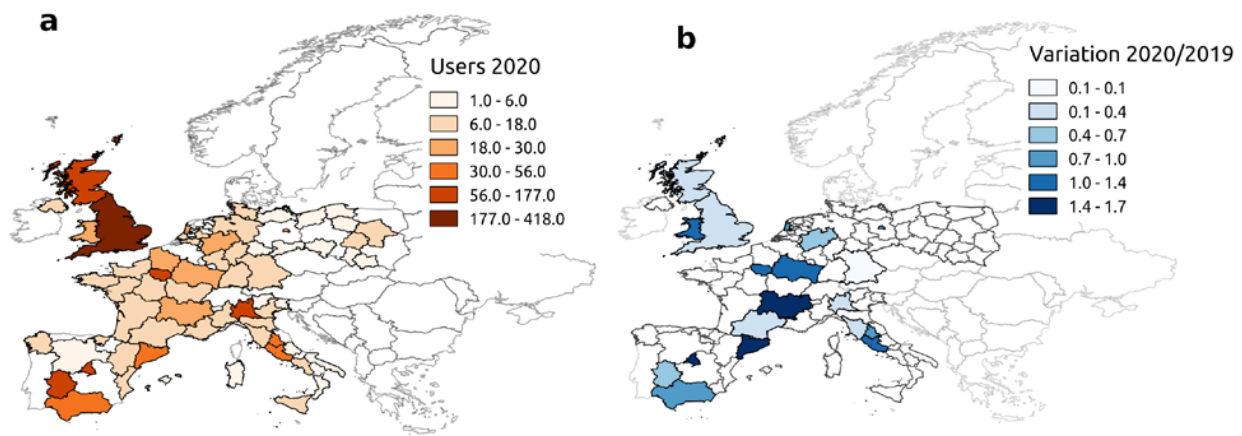

**Fig. 2. Geo-localization of pneumonia-related tweets posted across Europe since December 2019.** (a) Number of users discussing pneumonia between 15 December 2019 and 21 January 2020, after filtering out press releases and news accounts. (b) Relative variation in number of users discussing pneumonia between winter seasons 2019 and 2020.

To further check for robustness of findings, we also considered another common symptom that has been associated with COVID-19, i.e., dry cough [10]. Using the same procedure adopted for pneumonia (i.e., filtering out tweets induced by media exposure, comments of events reported in the news), we created a new data set containing all tweets mentioning dry cough, and computed the cumulative distribution of the number of such tweets. Figure 3a shows the anomalous increase in the number of these mentions during the weeks leading up to the peak in February 2020. We also computed the two-sample K-S test (Supplementary Fig. S4) and the two-sample A-D test (Supplementary Fig. S5) of the cumulative distributions related to dry cough over the two corresponding winter seasons (2018-2019 and 2019-2020). Figure 3b shows the geographic distribution of unique users that posted messages on dry cough between 1 December 2019 and 30 January 2020 (see also Supplementary Table S6). Findings are in agreement with the geographic distribution of users reporting on pneumonia in the same period: postings concerned with COVID-19-related symptoms preceded the official public announcements on local outbreaks, and were spatially concentrated in the areas that would subsequently become key infection hotspots.

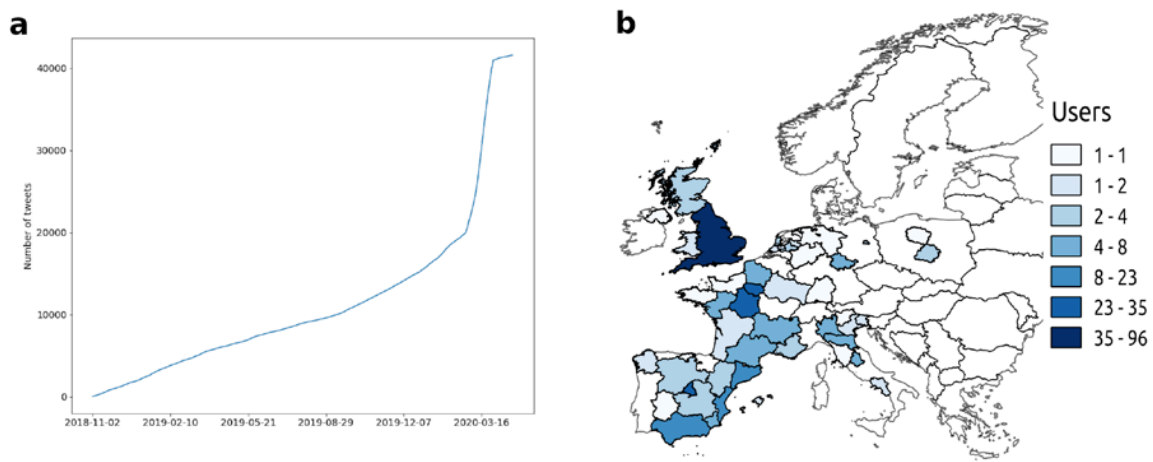

**Fig. 3. Anomalous evolution and geo-localization of tweets concerned with dry cough posted across Europe.** (a) Evolution of the cumulative number of dry cough-related tweets posted in 7 European countries since November 2018. (b) Geographic distribution of unique users discussing dry cough in the winter season between 1 December 2019 and 30 January 2020.

Finally, we controlled for a more general search term – "Coronavirus" – to ascertain whether messages broadly related to the epidemic, but not to personal medical symptoms, could uncover the effects of news exposure rather than genuine whistleblowing. We expected the geographical distribution of the users who posted such messages to differ from the spatial distribution of the actual hotspots of the epidemic. Notice that the term "COVID-19" could not be used as it was coined by WHO only on 21 January 2020, thus only towards the end of our focal 2019-2020 winter season. Figure 4a shows the geographic distribution of unique Twitter users citing Coronavirus between 1 December 2019 and 30 January 2020. The distribution of these users, whose interest in the infection threat was likely elicited by news exposure, is more uniform than the distribution of users posting on symptoms related to personal or personal-network experience. Further evidence on the impact of news exposure on collective attention can be found in the fact that the number of Twitter users citing Coronavirus from December 2019 to January 2020 correlates well with the population size of the European regions in which these users were located ($R^2$=0.968; Fig. 4b).

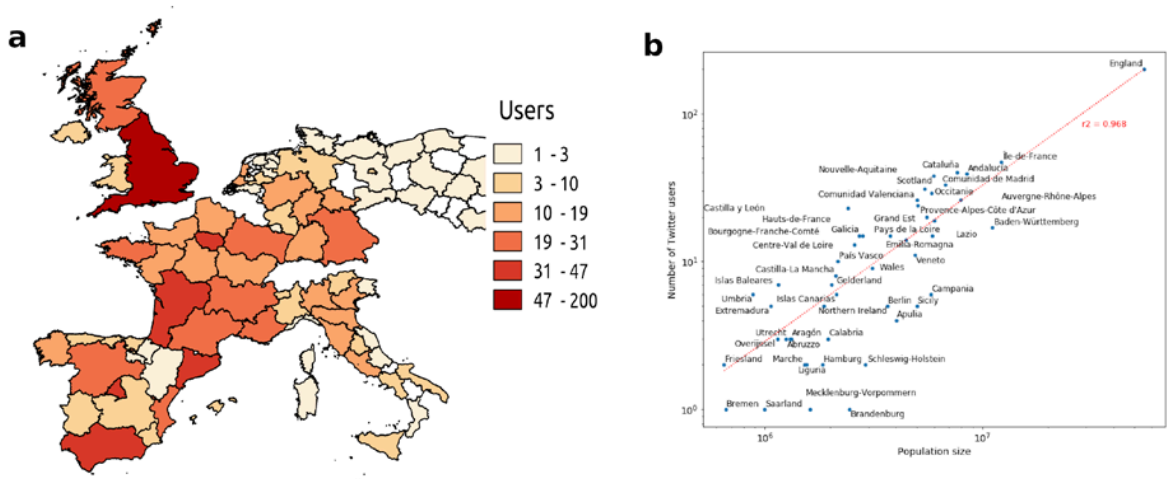

**Fig. 4. Geo-localization of tweets concerned with Coronavirus posted across Europe and relationship between number of users and population size.** (a) Geographic distribution of unique users discussing Coronavirus in the winter season between 1 December 2019 and 30 January 2020. (b) Scatter plot of the relationship between population size of European regions and number of unique users discussing Coronavirus in the same 2019-2020 winter season on a log-log scale. Coefficient of determination $R^2$ for the linear regression model shows a goodness of fit equal to 0.968. Data on population size was obtained from the official COVID-19 data set (source: John Hopkins University; https://github.com/CSSEGISandData/COVID-19).

**Discussion**

By leveraging social media, these findings offer the first clear accounting of how far behind many European countries were in detecting the virus. At the same time, the approach here outlined shows how governments, policy-makers and local authorities can obtain important contextual geo-localized information in real time for devising effective intervention policies throughout the whole epidemiological cycle, from the investigation and recognition phases of a pandemic up to the deceleration and preparation phases [14]. Recent studies have suggested how social media can help mitigate the psychological impact of COVID-19 on individuals, such as panic and depression among health practitioners and the general public [15]. In our work we showed how monitoring social media can also help public authorities to detect and geo-localize chains of contagion that would otherwise proliferate almost completely undetected for several weeks before the first death caused by a virus is announced. In turn, geo-localization of potential chains of infection could be effectively combined with data on atmospheric and environmental pollution, as part of an integrated early-intervention strategy for preventing epidemic spreads across geographical regions characterized by different exposure to environmental drivers of viral outbreaks [16].

Equally, social media can be used to mitigate the risk of a contagion resurgence in the phase 2 of a pandemic, when the restriction measures to counter the spread (e.g., social distancing) are progressively lifted. For example, in the current phase when many countries are still evaluating digital surveillance and contact-tracing solutions for large-scale adoption, using social media could help public health authorities to produce spatio-temporal density maps of infectious threats and ascertain which constraints can be relaxed and in which areas. This can help policy-makers and governments to differentiate and mitigate the social and economic consequences that restriction and lockdown measures introduced at a global scale might have in local regional areas [17].

A cautionary note is needed on the applicability of our study and its policy implications. Since the detection and geo-localization of potential viral outbreaks are based on suitable keywords clearly linked to well-known symptoms, our approach cannot be directly used for the forecasting of otherwise unknown diseases. Indeed the usage of the word "pneumonia" on Twitter could have served as a useful proper predictor only before pneumonia was publicly linked to COVID-19, and not at a time when news outlets and the public in general were already discussing it widely. Rather than a fully-fledged forecasting framework, our approach can be regarded as a nowcasting system for uncovering signals of (already existing) diseases that would otherwise remain hidden or be detected too late. Timely detection of such signals could indeed shed light on anomalous concentrations of diseases and, in general, help combat future pandemics.

Moreover, being able to promptly uncover early-warning signals can help identify hotspots of resurgent infections and help counter the threat of recurrent pandemic waves, especially in cases when the virus has not yet been eradicated and continues lingering in a population [18].

Our usage of social media across languages can pave the way towards a more integrated digital surveillance system that could, in principle, be managed by international health organizations at a global level, across geographical and institutional boundaries. This could help countries, within Europe and beyond, to better coordinate their healthcare, political, and socio-economic responses to initial outbreaks as well as the resurgence of subsequent waves of infection towards a more effective global strategy to address the threats of a pandemic. For example, using a unified digital surveillance system could help governments to better harmonize the timing and scale of country-level restriction measures affecting the activity and mobility of neighboring populations. In turn, devising a consistent set of domestic and cross-border responses could help secure the multilateral cooperation and integrated international effort needed for overcoming the global challenges of a pandemic.

In summary, social media hold promise for enhancing the effectiveness of public health surveillance, especially when combined with other novel data streams, such as Web search queries [19], participatory surveillance data [20], aggregated mobile phone data [21], and geospatial data from social contact tracing solutions [22]. Over the longer term, any integrated digital surveillance system set to monitor COVID-19 and beyond should be controlled by independent data protection and regulation authorities, and adhere to a clear set of privacy-preserving and data-sharing principles that do not jeopardize civil rights and other fundamental liberties.

**Methods**

Collected tweets were associated with the users' details by leveraging the Twitter API. We also collected information on the number of followers, friends, statuses and location of each user. The initial data set concerned with the winter seasons 2020-2019 and 2019-2018 included 573,298 unique users and a total of 891,195 unique tweets. From this data set we extracted a sample including tweets concerned with pneumonia and posted in the period between 15 December 2018 and 21 January 2019 and the period between 15 December 2019 and 21 January 2020. To conduct further robustness checks, we also extracted samples of tweets posted in all other corresponding winter seasons since 2014. From these tweets we selected those that were posted by users located in regions of the seven top-ranking countries according to number of speakers as percentage of the EU population (i.e., the United Kingdom, Germany, France, Italy, Spain, Poland, and the Netherlands) [13]. We used several geo-coders to cross-check the geographic coordinates of the users. In this way we could filter out the tweets from non-European English-speaking, Spanish-speaking and French-speaking countries. We used GIS methods to assign Twitter users to NUTS1 European regions (e.g., Lombardy, Île de France, Comunidad Valenciana).

To avoid overestimation of the number of tweets mentioning cases of pneumonia and mitigate bias in sample selection, we made the following adjustments: a) we removed all tweets (and the corresponding users) that cited news with a direct url; b) we considered only users with fewer than 2,000 followers to filter out the effects of press agencies and celebrities that usually have a large number of friends; c) we removed all remaining tweets that still contained the word "Coronavirus", "China", or "COVID". Applied to the period between December 2019 and January 2020, these steps reduced the number of tweets to a final sample of 4,765 and downsized the number of users to 2,716 by removing or mitigating the effects of COVID-19-related news that appeared up to 21 January 2020, when COVID-19 became a Class B notifiable disease [12].

We then identified the users that cited pneumonia in the selected European countries between 15 December 2019 and 21 January 2020, and compared with the total number of users that cited pneumonia in the same weeks of the previous year. To properly test the statistical significance of the change in number of pneumonia-related tweets, we used a methodology similar to the one applied to measure excess mortality in Europe [23]. In particular, we performed a two-sample Kolmogorov-Smirnov (K-S) test of the null hypothesis ($H_0$) that the cumulative distributions of number of tweets over the winter seasons 2018-2019 and 2019-2020 are the same against the alternative hypothesis ($H_1$) that the distributions are different. We focused on the following seasons: 15 December 2018 – 21 January 2019 (sample 1) and 15 December 2019 – 21 January 2020 (sample 2). When statistically significant, the difference between the cumulative distribution functions of the two samples is likely to indicate an increase in pneumonia-related tweets signaling a pneumonia diffusion pattern that differs from what would be expected from seasonal trends.

We proceeded as follows. We used moving windows of width between 50 and 70 days, and with daily frequency we computed, and averaged out, the K-S statistics over all these window widths. More specifically, for a given day $d_1$ and a given window width $w \in [50, 70]$:

1) we considered the two time intervals $W_1$ and $W_2$ in the two winter seasons of the same length $w$, i.e., starting and ending on the same days $d_1$ and $d_2$ in the seasons 2018-2019 and 2019-2020;

2) we computed the cumulative distributions of the number of tweets in $W_1$ and $W_2$;

3) we computed the K-S statistic (and the corresponding $p$-value);

4) we repeated steps 1-4 for each window width $w \in [50, 70]$, i.e., $w=50, ,51, 52, ...70$;

5) we averaged out the $p$-values over all window widths $w \in [50, 70]$;

6) we repeated steps 1-5 for every day $d$ of the winter seasons.

Thus, for every day of the focal winter season 2019-2020, we obtained an average of the K-S tests computed over moving windows of various widths. Figure 1b shows the $p$-values associated with the K-S test for the various European countries. Supplementary Table S1 reports, for each country, the dates (YYYY-MM-DD) between which the cumulative distributions differ.

To check for robustness, we also computed the Anderson-Darling (A-D) test using the same procedure as above (i.e., using moving windows of widths between 50 and 70 days and averaging out over these widths with a daily frequency). Supplementary Figure S1 shows results on the A-D test. These results are in agreement with the K-S test. Supplementary Table S2 reports, for each country, the dates (YYYY-MM-DD) between which the cumulative distributions differ (at the 0.05 level of significance). Notice that these time intervals are slightly more extended than the ones obtained with the K-S test, thus showing that the A-D test is more robust than the K-S one.

Using the same procedure, we also computed the K-S and A-D tests to compare the cumulative distribution of number of tweets citing pneumonia in the winter season 15 December 2019 – 21 January 2020 with the distributions for each of the corresponding seasonal periods since 2014. For each country, Supplementary Fig. S2 and Supplementary Fig. S3 show the average of the *p*-values obtained using each of the five winter seasons from 2014-2015 to 2018-2019, and computed from the K-S and A-D tests, respectively. Once again, the *p*-values related to each individual season are averages over moving window widths $w \epsilon [50, 70]$ computed with daily frequency. For each country, Supplementary Table S3 and Supplementary Table S4 report, respectively, the values of the K-S and A-D statistical tests (and corresponding *p*-values) comparing the 2019-2020 winter season with each of the preceding five seasons individually. Findings from all these tests are in agreement with previous results: with the exception of Germany, in the winter season 2019-2020 all selected European countries witnessed an excess posting of pneumonia-related tweets, with cumulative distributions that are statistically different from the ones obtained from each of the five preceding corresponding winter seasons.

Supplementary Table S5 shows the European regions associated with an excess of unique users discussing pneumonia between 15 December 2019 and 21 January 2020, after filtering out press releases and news accounts. The Table also highlights the regions that reported active local cases of COVID-19 in the initial period between 15 February and 7 March 2020.

We also computed the K-S test for the cumulative distributions related to dry cough in a similar way as before (i.e., average of window widths with daily frequency). In this case, the tests have been computed on cumulative data across languages/countries. Notice that the total number of mentions (tweets) of dry cough across 7 languages in the 10 years before COVID-19 took place is less than 1,000 per year, thus making the curve fragmented and difficult to interpret. Summing up all the data across languages would produce a better signal, and also enable us to filter out local phenomena and smooth out statistical variability.

Supplementary Figure S4 shows results on the K-S test of the null hypothesis that the cumulative distributions of number of tweets citing dry cough over the two corresponding winter seasonal periods (2018-2019 and 2019-2020) are the same against the alternative hypothesis that the distributions are statistically different. As with tweets mentioning pneumonia, for robustness check we also computed the A-D test on the cumulative distributions of number of tweets citing dry cough. Supplementary Figure S5 shows the results of the A-D test.

Supplementary Table S6 reports the number of unique Twitter users per European region that posted an anomalous volume of messages mentioning dry cough between 1 December 2019 and 30 January 2020, that is before the first announcements of the COVID-19 epidemic were officially made in Europe.

# Supplementary Information

Early warnings of COVID-19 outbreaks across Europe from social media?

Milena Lopreite, Pietro Panzarasa, Michelangelo Puliga, Massimo Riccaboni

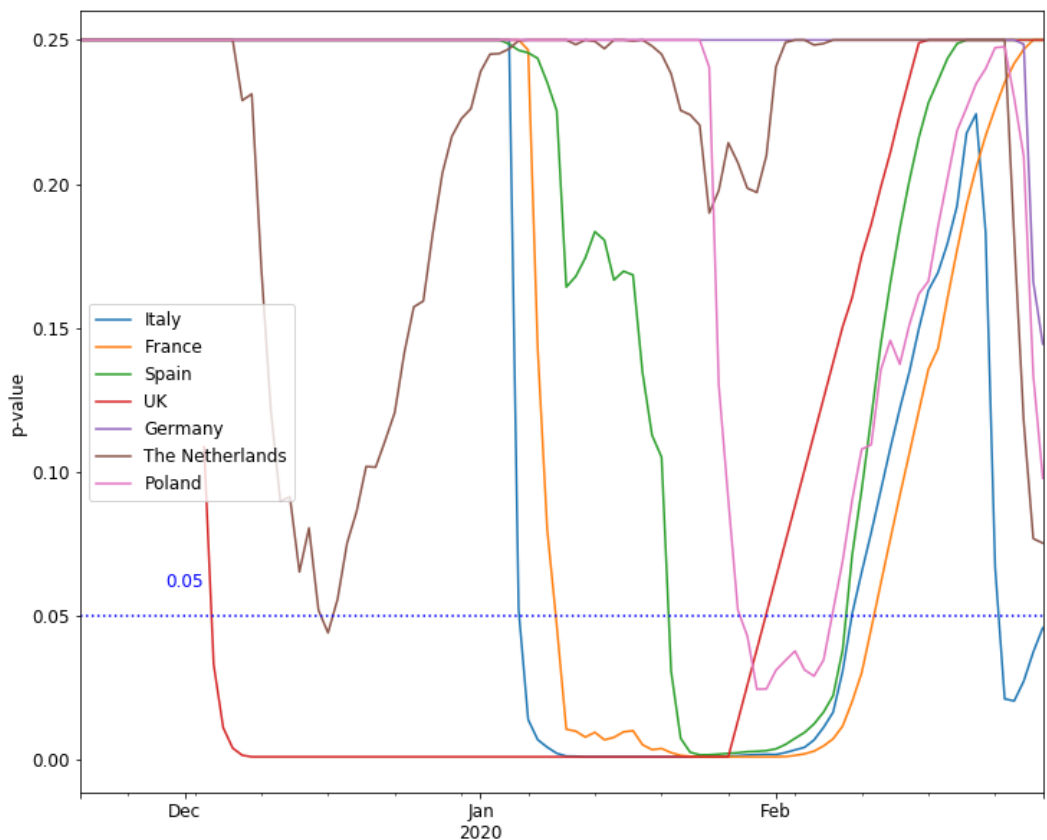

**Fig. S1 Two-sample Anderson-Darling test of the difference between cumulative distributions of number of tweets citing pneumonia and posted in two corresponding winter seasonal periods (2018-2019 and 2019-2020) for each of the 7 European countries**. The graph reports the average *p*-values over moving window widths $w \in [50, 70]$ computed with daily frequency.

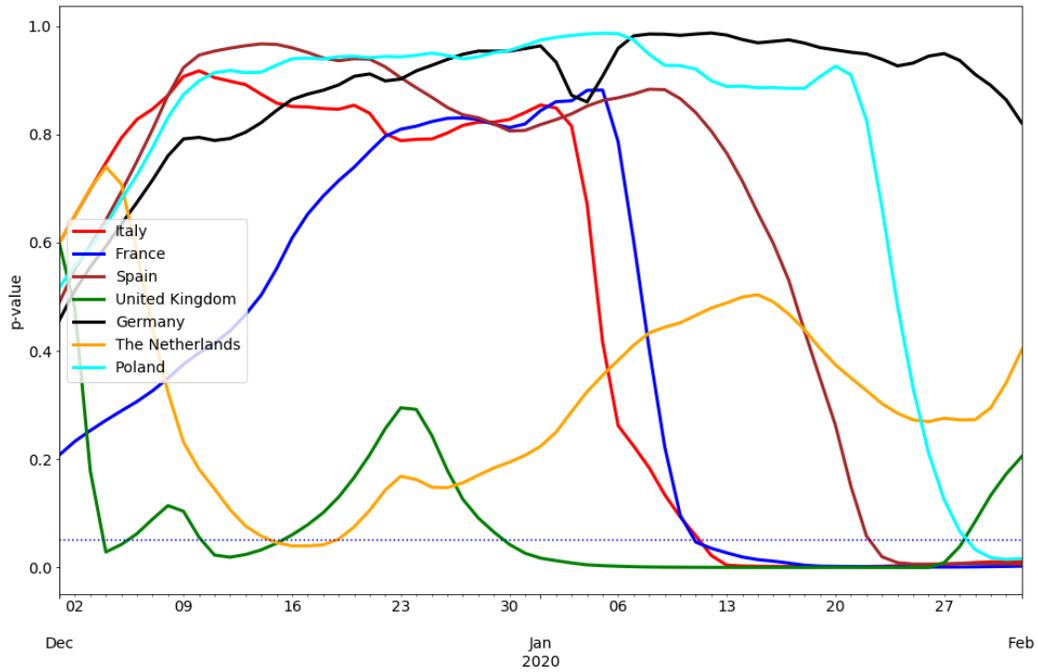

**Fig. S2 Two-sample Kolmogorov-Smirnov test of the difference between cumulative distributions of number of tweets citing pneumonia and posted in the winter seasonal period 2019-2020 and in the corresponding five preceding periods since 2014-2015 for each of the 7 European countries**. The graph reports the average of the $p$-values obtained using each of the five preceding winter seasons. In turn, the $p$-values related to each individual season are averages over moving window widths $w \, \epsilon \, [50, 70]$ computed with daily frequency.

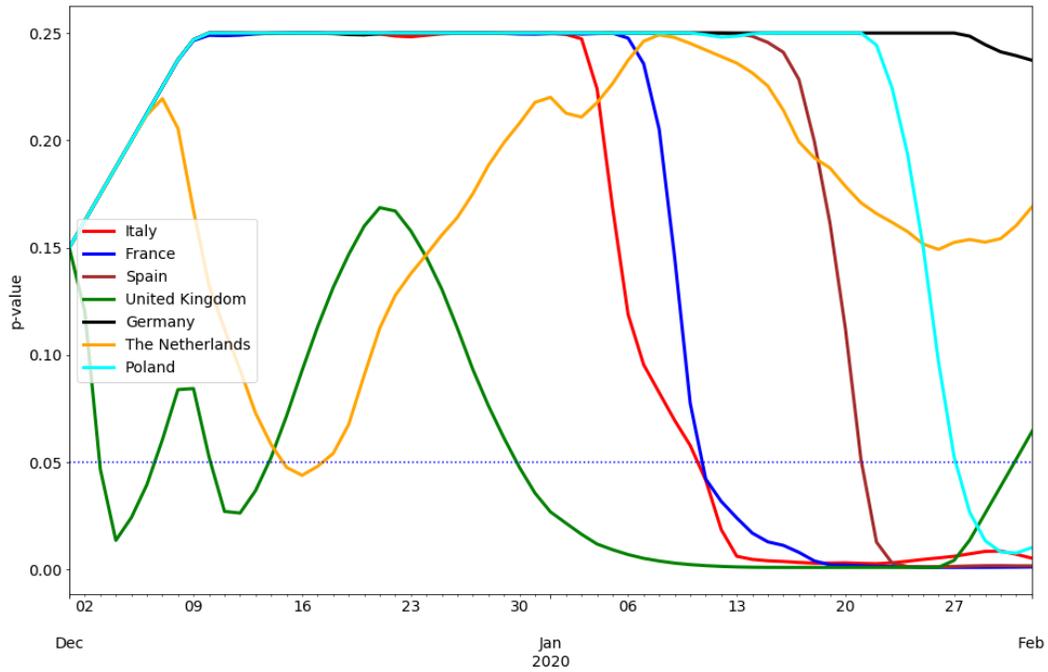

**Fig. S3 Two-sample Anderson-Darling test of the difference between cumulative distributions of number of tweets citing pneumonia and posted in the winter seasonal period 2019-2020 and in the corresponding five preceding periods since 2014-2015 for each of the 7 European countries**. The graph reports the average of the $p$-values obtained using each of the five preceding winter seasons. In turn, the $p$-values related to each individual season are averages over moving window widths $w \, \epsilon \, [50, 70]$ computed with daily frequency.

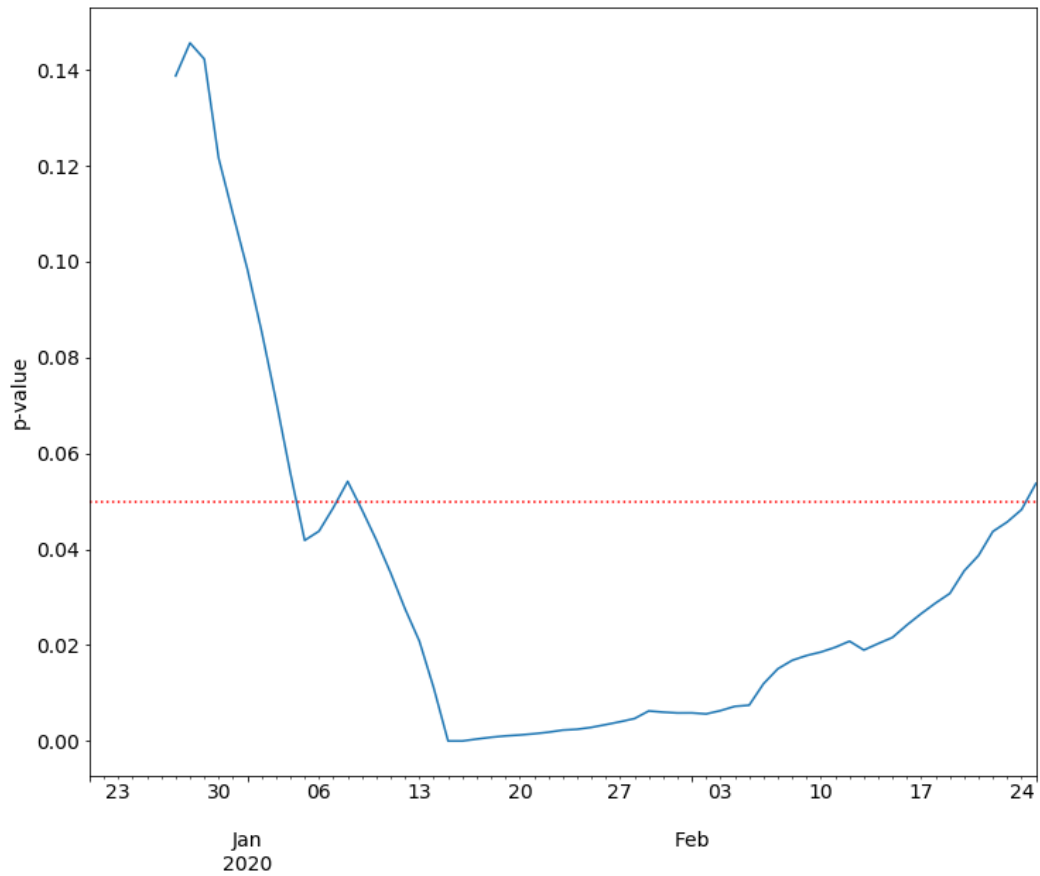

**Fig. S4 Two-sample Kolmogorov-Smirnov test of the difference between cumulative distributions of number of tweets citing dry cough and posted in two corresponding winter seasonal periods (2018-2019 and 2019-2020) in 7 European countries**. The graph reports the average $p$-values over moving window widths $w \,\epsilon\, [50, 70]$ computed with daily frequency.

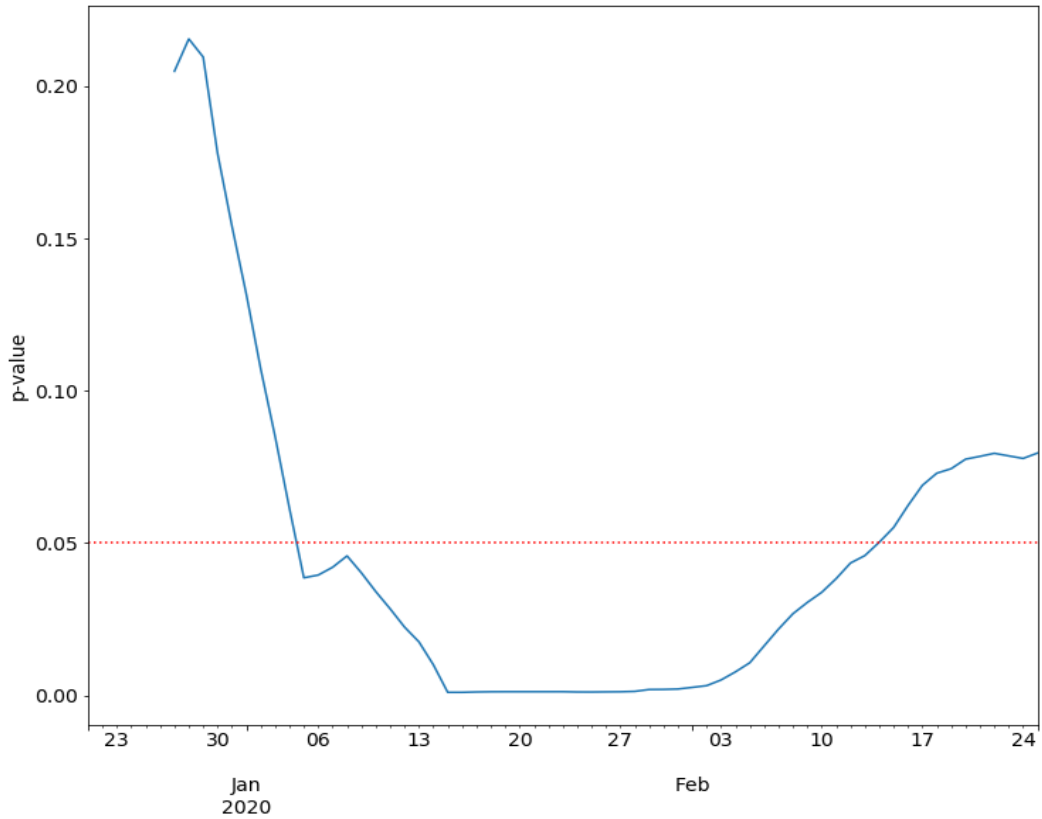

**Fig. S5 Two-sample Anderson-Darling test of the difference between cumulative distributions of number of tweets citing dry cough and posted in two corresponding winter seasonal periods (2018-2019 and 2019-2020) in 7 European countries**. The graph reports the average *p*-values over moving window widths $w \ \epsilon \ [50, 70]$ computed with daily frequency.

| Countries | Periods |
|---|---|
| Italy | 2020/01/06 - 2020/02/07 |
| France | 2020/01/10 - 2020/02/08 |
| Spain | 2020/01/22 - 2020/02/06 |
| UK | 2019/12/05 - 2020/01/28 |
| The Netherlands | 2019/12/15 - 2019/12/18 |
| Poland | 2020/02/05 - 2020/02/06 |

**Table S1 Countries and time periods in the 2019-2020 winter season characterized by excess posting on pneumonia according to the Kolmogorov-Smirnov tests.** In these periods the cumulative distributions of pneumonia-related tweets differ from the distributions in the corresponding time periods in the 2018-2019 season (K-S tests at the 0.05 level of significance).

| Country | Period |
|---|---|
| Italy | 2020/01/06 – 2020/02/29 |
| France | 2020/01/09 – 2020/02/11 |
| Spain | 2020/01/21 – 2020/02/08 |
| UK | 2019/12/04 – 2020/01/30 |
| The Netherlands | 2019/12/16 – 2019/12/16 |
| Poland | 2020/01/29 – 2020/02/06 |

**Table S2 Countries and time periods in the 2019-2020 winter season characterized by excess posting on pneumonia according to Anderson-Darling tests.** These are the periods in which the cumulative distributions of dry cough-related tweets differ from the distributions in the corresponding time periods in the 2018-2019 season (A-D tests at the 0.05 level of significance).

| Country | 2014-2015 | 2015-2016 | 2016-2017 | 2017-2018 | 2018-2019 |
|---|---|---|---|---|---|
| **France** | 0.36066 | 0.39344 | 0.36066 | 0.44262 | 0.40984 |
| | (0.00064) | (0.00013) | (0.00064) | (0.00001) | (0.00006) |
| **Germany** | 0.09836 | 0.06530 | 0.08197 | 0.14754 | 0.06557 |
| | (0.93268) | (0.99742) | (0.98783) | (0.52379) | (0.99956) |
| **Italy** | 0.34426 | 0.39344 | 0.34426 | 0.34426 | 0.45902 |
| | (0.00133) | (0.00013) | (0.00133) | (0.00133) | (0.00000) |
| **The Netherlands** | 0.26230 | 0.18033 | 0.27869 | 0.21173 | 0.22812 |
| | (0.02967) | (0.27603) | (0.01712) | (0.11129) | (0.07130) |
| **Poland** | 0.37324 | 0.36548 | 0.31064 | 0.27581 | 0.26202 |
| | (0.00036) | (0.00075) | (0.00452) | (0.01716) | (0.02566) |
| **Spain** | 0.34426 | 0.32787 | 0.32787 | 0.32787 | 0.31148 |
| | (0.00133) | (0.00266) | (0.00266) | (0.00266) | (0.00512) |
| **UK** | 0.37705 | 0.37705 | 0.42623 | 0.37705 | 0.36066 |
| | (0.00030) | (0.00030) | (0.00002) | (0.00030) | (0.00064) |

**Table S3 Two-sample Kolmogorov-Smirnov test of the difference between cumulative distributions of number of tweets citing pneumonia posted in the winter seasonal period 2019-2020 and in each of the corresponding five periods since 2014-2015 for each of the 7 European countries**. For each country and winter seasonal period, the table reports the values of the statistical test (corresponding *p*-values are shown in parentheses).

| Country | 2014-2015 | 2015-2016 | 2016-2017 | 2017-2018 | 2018-2019 |
|---|---|---|---|---|---|
| **France** | 9.39622 | 10.05256 | 7.00257 | 14.51323 | 9.64127 |
| | (0.00100) | 0.00100) | (0.00100) | (0.00100) | (0.00100) |
| **Germany** | -0.82499 | -1.13954 | -1.05783 | 0.62221 | -1.09790 |
| | (0.25000) | (0.25000) | (0.25000) | (0.18304) | (0.25000) |
| **Italy** | 8.17272 | 11.00139 | 5.74394 | 5.05182 | 12.30459 |
| | (0.00100) | (0.00100) | (0.00189) | (0.00332) | (0.00100) |
| **The Netherlands** | 2.77755 | 1.01411 | 2.39161 | 0.52939 | 1.64154 |
| | (0.02381) | (0.12474) | (0.03388) | (0.20061) | (0.06830) |
| **Poland** | 8.38603 | 6.24365 | 4.13372 | 3.03733 | 3.73690 |
| | (0.00100) | (0.00128) | (0.00719) | (0.01883) | (0.01014) |
| **Spain** | 7.95646 | 7.28634 | 7.26826 | 6.87016 | 6.12392 |
| | (0.00100) | (0.00100) | (0.00100) | (0.00100) | (0.00140) |
| **UK** | 5.28041 | 5.36497 | 7.49829 | 5.45749 | 4.34225 |
| | (0.00276) | (0.00257) | (0.00100) | (0.00239) | (0.00602) |

**Table S4 Two-sample Anderson-Darling test of the difference between cumulative distributions of number of tweets citing pneumonia posted in the winter seasonal period 2019-2020 and in each of the corresponding five periods since 2014-2015 for each of the 7 European countries**. For each country and winter seasonal period, the table reports the values of the statistical test (corresponding $p$-values are shown in parentheses).

| Country/Region | Users 2020 | Users 2019 | Relative variation 2020-2019 | Absolute variation 2020-2019 |
|---|---|---|---|---|
| **Germany** Total number of tweets = 452 | | | | |
| Rheinland-Pfalz | 14 | 4 | 2.50 | 10 |
| Hessen | 28 | 9 | 2.11 | 19 |
| Baden-Württemberg* | 27 | 11 | 1.45 | 16 |
| Nordrhein-Westfalen* | 46 | 19 | 1.42 | 27 |
| Schleswig-Holstein | 14 | 6 | 1.33 | 8 |
| Hamburg | 18 | 8 | 1.25 | 10 |
| Berlin | 56 | 30 | 0.87 | 26 |
| Bayern* | 26 | 20 | 0.30 | 6 |
| Niedersachsen | 14 | 12 | 0.17 | 2 |
| Total number of users | 243 | 119 | 1.04 | 124 |
| **Spain** Total number of tweets = 2,245 | | | | |
| Castilla-La Mancha* | 11 | 1 | 10.00 | 10 |
| Comunidad de Madrid* | 203 | 52 | 2.90 | 151 |
| Cataluña | 122 | 34 | 2.59 | 88 |
| Aragón | 11 | 4 | 1.75 | 7 |
| Extremadura | 158 | 68 | 1.32 | 90 |
| Islas Canarias | 13 | 6 | 1.17 | 7 |
| Andalucía | 83 | 42 | 0.98 | 41 |
| Galicia | 15 | 8 | 0.88 | 7 |
| Comunidad Valenciana | 38 | 24 | 0.58 | 14 |
| País Vasco | 11 | 7 | 0.57 | 4 |
| Total number of users | 665 | 246 | 1.70 | 419 |

**France**
Total number of tweets = 2,112

| | | | | |
|---|---|---|---|---|
| Provence-Alpes-Côte d'Azur | 57 | 11 | 4.18 | 46 |
| Bretagne | 24 | 6 | 3.00 | 18 |
| Centre-Val de Loire | 30 | 8 | 2.75 | 22 |
| Grand Est* | 54 | 15 | 2.60 | 39 |
| Auvergne-Rhône-Alpes | 67 | 19 | 2.53 | 48 |
| Île-de-France* | 361 | 105 | 2.44 | 256 |

| | | | |
|---|---|---|---|
| Normandie | 32 | 10 | 2.20 | 22 |
| Nouvelle-Aquitaine | 42 | 14 | 2.00 | 28 |
| Hauts-de-France | 49 | 20 | 1.45 | 29 |
| Pays de la Loire | 36 | 15 | 1.40 | 21 |
| Occitanie | 43 | 19 | 1.26 | 24 |
| Bourgogne-Franche-Comté | 21 | 12 | 0.75 | 9 |
| Total number of users | 816 | 254 | 2.21 | 562 |

**Italy**
Total number of tweets = 1,097

| | | | |
|---|---|---|---|
| Friuli-Venezia Giulia | 11 | 2 | 4.50 | 9 |
| Piemonte* | 20 | 7 | 1.86 | 13 |
| Emilia-Romagna* | 19 | 7 | 1.71 | 12 |
| Umbria | 87 | 44 | 0.98 | 43 |
| Lazio | 61 | 32 | 0.91 | 29 |
| Veneto | 20 | 12 | 0.67 | 8 |
| Campania | 16 | 10 | 0.60 | 6 |
| Sicily | 19 | 12 | 0.58 | 7 |
| Toscana | 23 | 16 | 0.44 | 7 |
| Lombardia* | 201 | 151 | 0.33 | 50 |
| Total number of users | 477 | 293 | 0.63 | 184 |

**The Netherlands**
Total number of tweets = 380

| | | | | |
|---|---|---|---|---|
| Noord-Brabant* | 16 | 8 | 1.00 | 8 |
| Zuid-Holland | 33 | 17 | 0.94 | 16 |
| Gelderland | 17 | 9 | 0.89 | 8 |
| Noord-Holland | 52 | 30 | 0.73 | 22 |
| Total number of users | 118 | 64 | 0.84 | 54 |

**Poland**
Total number of tweets = 244

| | | | | |
|---|---|---|---|---|
| Mazowieckie* | 25 | 10 | 1.50 | 15 |
| Łódzkie | 31 | 17 | 0.82 | 14 |
| Total number of users | 56 | 27 | 1.07 | 29 |

**United Kingdom**
Total number of tweets = 4,451

| | | | |
|---|---|---|---|
| England* | 1,462 | 484 | 2.02 | 978 |
| Wales* | 66 | 22 | 2.00 | 44 |
| Northern Ireland | 36 | 14 | 1.57 | 22 |
| Scotland | 192 | 83 | 1.31 | 109 |
| Total number of users | 1,756 | 603 | 1.91 | 1,153 |

**Table S5 European regions associated with an excess number of unique users discussing pneumonia, after filtering out press releases and news accounts**. Highlighted in red are the regions that, based on the Wikipedia pages that summarize the statistics on COVID-19 per each country, reported active local cases of COVID-19 in the initial period between 15 February and 7 March 2020. Marked with a star (*) are the regions with the highest number of COVID-19 cases per capita (Source: Wikipedia, 20 April 2020).

| European region | Number of unique users |
|---|---|
| England | 96 |
| Comunidad de Madrid | 35 |
| Île-de-France | 27 |
| Centre-Val de Loire | 26 |
| Andalucía | 23 |
| Cataluña | 17 |
| Comunidad Valenciana | 14 |
| Hauts-de-France | 8 |
| Lombardia | 8 |
| Umbria | 7 |
| Thüringen | 6 |
| Pays de la Loire | 6 |
| Auvergne-Rhône-Alpes | 6 |
| Occitanie | 6 |
| Región de Murcia | 5 |
| Emilia-Romagna | 5 |
| Provence-Alpes-Côte d'Azur | 4 |
| Castilla-La Mancha | 4 |
| Scotland | 3 |
| Noord-Holland | 3 |
| Aragón | 3 |
| Łódzkie | 3 |
| Castilla y León | 3 |
| Gelderland | 3 |
| Berlin | 3 |
| Islas Baleares | 2 |
| Wales | 2 |
| Overijssel | 2 |
| Nouvelle-Aquitaine | 2 |
| Utrecht | 2 |
| Veneto | 2 |
| Friuli-Venezia Giulia | 2 |
| Zuid-Holland | 2 |
| Galicia | 2 |
| Grand Est | 2 |
| Campania | 2 |
| Bretagne | 1 |
| Baden-Württemberg | 1 |
| Trentino-Alto Adige | 1 |
| Toscana | 1 |
| Groningen | 1 |
| Cantabria | 1 |
| Hamburg | 1 |
| Extremadura | 1 |
| Northern Ireland | 1 |
| Normandie | 1 |
| Nordrhein-Westfalen | 1 |
| Niedersachsen | 1 |
| Kujawsko-Pomorskie | 1 |
| Ceuta y Melilla | 1 |

**Table S6 European regions and number of unique users discussing dry cough between 1 December 2019 and 30 January 2020.** Usual adjustments have been made to filter out messages related to press releases and news account.